\documentclass[aps,prb,reprint,superscriptaddress]{revtex4-2}

\usepackage{graphicx}
\usepackage{dcolumn}
\usepackage{bm}
\usepackage{amsmath}
\usepackage{amssymb}
\usepackage{hyperref}
\begin{document}

\title{Entropy-Driven Structural Phase Transition in \texorpdfstring{Nb$_3$Cl$_8$}{Nb3Cl8} via DFT and an Effective Model}

\author{Chenjie Zhu}
\affiliation{Beijing National Laboratory for Condensed Matter Physics and Institute of Physics, Chinese Academy of Sciences, Beijing 100190, China}

\author{Shuai Zhang}
\affiliation{Institute of Theoretical Physics, Chinese Academy of Sciences, Beijing 100190, China}
\author{Zhong Fang}
\affiliation{Beijing National Laboratory for Condensed Matter Physics and Institute of Physics, Chinese Academy of Sciences, Beijing 100190, China}
\author{Zhijun Wang}
\affiliation{Beijing National Laboratory for Condensed Matter Physics and Institute of Physics, Chinese Academy of Sciences, Beijing 100190, China}
\author{Quansheng Wu}
\affiliation{Beijing National Laboratory for Condensed Matter Physics and Institute of Physics, Chinese Academy of Sciences, Beijing 100190, China}

\author{Hongming Weng}
\email{hmweng@iphy.ac.cn}
\affiliation{Beijing National Laboratory for Condensed Matter Physics and Institute of Physics, Chinese Academy of Sciences, Beijing 100190, China}

\date{\today}

\begin{abstract}
As a prototypical flat-band cluster Mott insulator on an effective triangular lattice, $\mathrm{Nb}_3\mathrm{Cl}_8$ is a potential candidate for hosting a quantum spin liquid (QSL) state. Nevertheless, a first-order structural phase transition around $90\,\mathrm{K}$ transforms the high-temperature paramagnetic $\alpha$ phase into the low-temperature nonmagnetic $\beta$ phase, suppressing the candidate QSL regime of the $\alpha$ phase. To clarify the microscopic origin of this transition, we combine first-principles calculations with an extended Hubbard model to construct a unified free-energy framework. This framework reveals that the transition is jointly driven by phonon and spin entropy: the $\alpha$ phase is stabilized by softer phonons and larger paramagnetic spin entropy, whereas the $\beta$ phase is favored by interlayer dimerization, which hardens the phonons and quenches the spin entropy through singlet formation. Furthermore, by evaluating the pressure-dependent generalized enthalpy, we provide a thermodynamic explanation for the suppression of the transition under c-axis uniaxial pressure, where stabilizing the $\alpha$ phase may allow the candidate QSL regime of the $\alpha$ phase to be explored at low temperatures. 
\end{abstract}

\maketitle

\section{Introduction}

Strongly correlated electron systems host emergent phenomena such as Mott insulating behavior\cite{MIT}, high-temperature superconductivity\cite{HTCSC,HTCSC2}, and charge-density-wave order\cite{CDW}. Identifying their microscopic origins remains a central challenge because orbital, spin, charge, and lattice degrees of freedom are often strongly intertwined.

A canonical example is 1T-$\mathrm{TaS}_2$, where the commensurate superlattice within its charge-density-wave phase produces strong lattice distortions, localizing electrons onto David-star clusters and driving the resulting half-filled narrow band into a cluster Mott insulating state\cite{1T-TaS2_1,1T-TaS2_2}. Even in this case, however, the interplay between electron correlations and interlayer coupling complicates the origin of the insulating behavior\cite{1T-TaS2_3}.

It is therefore useful to identify material platforms with simpler structures and more transparent microscopic mechanisms, which can serve as benchmarks for more complicated correlated systems.

The layered van der Waals breathing-kagome family $\mathrm{Nb}_3\mathrm{X}_8$ ($\mathrm{X = Cl, Br, I}$) has recently emerged as a promising platform for cluster Mott physics. Strong $\mathrm{Nb}-\mathrm{Nb}$ bonding trimerizes neighboring Nb atoms into $\mathrm{Nb}_3\mathrm{X}_{13}$ clusters, which form a triangular sublattice within each layer. Each $[\mathrm{Nb}_3]^{8+}$ cluster contributes seven 4d electrons, leaving one unpaired electron in the $2a_1$ molecular orbital\cite{Tunable,Discovery}. In monolayer $\mathrm{Nb}_3\mathrm{Cl}_8$, weak intercluster hopping produces an isolated flat $2a_1$ band near the Fermi level, realizing a half-filled cluster Mott insulator well described by a single-band Hubbard model\cite{Discovery,Calc2023,MonolayerMottPRL,ThinFlakeMottNSR}.

Bulk $\mathrm{Nb}_3\mathrm{Cl}_8$ exhibits richer behavior. At high temperature, it behaves as a paramagnetic cluster Mott insulator with spin-$1/2$ local moments. Below about $90\ \mathrm{K}$\cite{Rearrangement}, it undergoes a first-order structural phase transition into a low-temperature nonmagnetic phase, accompanied by a sharp drop in magnetic susceptibility\cite{Disproportionation,Rearrangement,Tunable,observation,NMR_CPL_2025}. Electrical transport measurements show that both bulk phases are insulating\cite{conductivity}. This phase transition primarily involves an interlayer rearrangement that modifies the coupling between neighboring layers. Recent momentum-resolved High-Resolution Electron Energy Loss Spectroscopy (HREELS) measurements further show that the excitation spectrum is reconstructed across the transition: a quasi-2D linearly dispersing exciton in the $\alpha$ phase evolves into split excitonic bands with three-dimensional parabolic dispersion in the $\beta$ phase, consistent with enhanced interlayer coupling and bonding-antibonding splitting\cite{ExcitonPRL2026}.

The microscopic origin of the vanishing magnetic susceptibility in the $\beta$ phase remains under debate. Several closely related structural models, including $\mathrm{C2/m}$\cite{Rearrangement}, $\mathrm{R3}$\cite{Disproportionation}, and $\mathrm{R\bar{3}m}$\cite{suppression}, have been proposed; they share the same interlayer-rearranged stacking motif but differ in subtle symmetry lowering and internal atomic positions. One explanation attributes the moment quenching to charge disproportionation, $2[\mathrm{Nb}_3]^{8+} \rightarrow [\mathrm{Nb}_3]^{7+} + [\mathrm{Nb}_3]^{9+}$, inferred from slight layer-dependent differences in intratrimer $\mathrm{Nb}-\mathrm{Nb}$ bond lengths in the $\mathrm{R3}$ structure, where a singlet is proposed to form on the expanded trimer\cite{Disproportionation}. However, the structural assignment of the $\beta$ phase remains unsettled. The required bond length differences are absent in the $\mathrm{C2/m}$ and $\mathrm{R\bar{3}m}$ models, and theoretical calculations indicate that interlayer $\mathrm{Nb}-\mathrm{Nb}$ bond length differences alone are insufficient to induce charge ordering\cite{Hubbarddimer}.

Another widely discussed scenario is the formation of interlayer dimer singlets. In this picture, the strong interlayer coupling in the $\beta$ phase generates effective antiferromagnetic correlations between neighboring clusters and stabilizes singlets on interlayer dimers\cite{Rearrangement}. Although these studies clarify the contrast in interlayer coupling between the $\alpha$ and $\beta$ phases, the thermodynamic driving force of the first-order transition has remained unclear. The mechanisms by which thinning\cite{pressure,suppression}, c-axis uniaxial pressure\cite{pressure}, and powdering\cite{Disproportionation,pressure} suppress the structural transition are likewise unresolved. Since the transition masks the low-temperature physics of the $\alpha$ phase, a promising quantum spin liquid (QSL) candidate on a frustrated triangular lattice, suppressing it may provide an experimental route to this state\cite{pressure,MuSR_NMR_2025}.

Recent studies emphasize that phase transitions in correlated materials are often governed by the interplay of multiple thermodynamic degrees of freedom—such as phonon entropy in $\mathrm{VO_2}$\cite{VO2_2,Souvatzis2008}, coupled spin-lattice effects in $\mathrm{FeRh}$\cite{FeRh2020} and $\mathrm{YNiO_3}$\cite{YNiO3Thermo}, or stationary DFT + DMFT energetics in $\mathrm{Ce}$\cite{HauleBirol2015}. These examples motivate our free-energy approach for $\mathrm{Nb_3Cl_8}$, explicitly separating lattice, spin, and correlation-related contributions rather than assuming a single driving force.

In this work, we investigate the first-order structural phase transition in $\mathrm{Nb}_3\mathrm{Cl}_8$ by combining first-principles calculations with a low-energy model Hamiltonian. We construct a free-energy framework that separates lattice and spin contributions, allowing us to resolve the thermodynamic competition between the entropy-driven $\alpha$ phase and the interlayer-dimerized, spin-singlet $\beta$ phase. This free-energy framework successfully reproduces a semiquantitative transition temperature, a structural barrier consistent with the observed thermal hysteresis, and a spin-entropy change that agrees with the macroscopic measurements\cite{Rearrangement}. We further show that c-axis uniaxial pressure suppresses the transition by stabilizing the $\alpha$ phase in the generalized-enthalpy landscape, whereas c-axis tension is expected to favor the $\beta$ phase. These results establish a unified thermodynamic picture of the structural transition in $\mathrm{Nb}_3\mathrm{Cl}_8$ and its pressure response.

\section{methods}

At zero external stress, the relative stability of the $\alpha$ and $\beta$ phases is determined by their Helmholtz free energies. We evaluate this quantity within an approximate framework that combines the non-spin-polarized DFT total energy $E_{\mathrm{DFT}}$, the harmonic phonon free energy $F_{\mathrm{phonon}}(T)$, and the free-energy contribution $F_{\mathrm{Hub}}(T)$ from an extended Hubbard model constructed in the $2a_1$ correlated subspace\cite{Calc2023}. A double-counting correction $E_{\mathrm{DC}}$ is subtracted to remove from the DFT reference the low-energy $2a_1$ contribution treated explicitly by the model Hamiltonian:
\begin{equation}
\label{eq:free_energy_total}
F_\mathrm{tot}(T) = E_{\mathrm{DFT}} + F_{\mathrm{phonon}}(T) - E_{\mathrm{DC}} + F_{\mathrm{Hub}}(T).
\end{equation}

\subsection{DFT electronic structure}

Electronic structure calculations were performed with VASP (Vienna Ab initio Simulation Package)\cite{vasp1,vasp2} to obtain the relaxed structures, ground-state total energies, and electronic band structures. The calculations presented in the results were performed using the nonlocal optB88-vdW functional, while the Perdew-Burke-Ernzerhof (PBE) generalized-gradient approximation was used as a robustness check. For both the $\alpha$ and $\beta$ phases, we used a $4 \times 4 \times 4$ $\mathbf{k}$-point mesh and a plane-wave cutoff of 450 eV. Structural relaxations were carried out with $\mathrm{EDIFFG} = -1 \times 10^{-4}\,\text{eV/\AA{}}$ and $\mathrm{ISIF} = 3$, allowing both atomic positions and lattice parameters to relax while preserving the crystal symmetry.

To estimate the energy barrier associated with the structural transformation between the relaxed $\alpha$ and $\beta$ phases, we construct a linear interpolation path connecting the two optimized structures. Along this path, both the lattice parameters and internal atomic coordinates are uniformly interpolated, yielding 10 structures in total, including the two relaxed endpoints, and no additional structural relaxation is performed for the intermediate configurations.

\subsection{Phonon thermodynamics}

Phonon properties were computed within the harmonic approximation using density functional perturbation theory (DFPT) implemented in VASP, with Phonopy used for pre-processing and post-processing\cite{phonopy1,phonopy2}. For each relaxed crystal structure, DFPT was used to obtain the harmonic force constants, from which the phonon frequencies $\omega_{\mathbf{q}\nu}$ were obtained by diagonalizing the corresponding dynamical matrices. The DFPT force constants were calculated using a $2 \times 2 \times 2$ supercell and a $1 \times 1 \times 1$ q-point mesh. The phonon density of states and the thermodynamic functions were then evaluated by sampling the phonon frequencies on a dense $31 \times 31 \times 31$ q-point mesh.

Within the harmonic approximation, the phonon Helmholtz free energy was evaluated by integrating the thermodynamic contributions of individual modes:
\begin{equation}
    \label{eq:F_phonon}
    F_{\mathrm{phonon}}(T) = \sum_{\mathbf{q}\nu} \left[ \frac{1}{2}\hbar\omega_{\mathbf{q}\nu} + \frac{1}{\beta}\ln\left(1-e^{-\beta\hbar\omega_{\mathbf{q}\nu}}\right) \right].
\end{equation}
The phonon internal energy can be obtained from $U_{\mathrm{phonon}} = \partial(\beta F_{\mathrm{phonon}})/\partial\beta$ (where $\beta = 1/k_B T$). Here $E_{\mathrm{zpe}} = U_{\mathrm{phonon}}(0)$ is the zero-point energy. The entropy can be obtained from $S_{\mathrm{phonon}} = -\partial F_{\mathrm{phonon}}/\partial T$. In this treatment, the phonon frequencies are those of the relaxed zero-temperature structures; temperature-induced phonon renormalization and anharmonic effects are not included.

\subsection{Hubbard model and double-counting correction}

Given that standard non-spin-polarized DFT cannot describe the Mott physics in the half-filled $2a_1$ manifold, we treat this low-energy subspace explicitly using an extended Hubbard model. The model is constructed in the Wannier basis of the cluster-centered $2a_1$ orbital, with one correlated orbital per $[\mathrm{Nb}_3]^{8+}$ cluster and the on-site energy chosen as the reference. The low-energy Hamiltonian reads
\begin{equation}
\begin{aligned}
    \label{eq:H_Hub}
    \hat H_{\mathrm{Hub}} ={}&
    \sum_{\langle i,j\rangle,\sigma}
    \left(t_{ij}\hat c_{i\sigma}^{\dagger}\hat c_{j\sigma}+\mathrm{H.c.}\right) 
    +\sum_i U_i \hat n_{i\uparrow}\hat n_{i\downarrow} \\
    &+\sum_{\langle i,j\rangle}V_{ij}\hat n_i\hat n_j .
\end{aligned}
\end{equation}
Here $t_{ij}$ denotes the Wannier hopping amplitude, while $U_i$ and $V_{ij}$ are the screened on-site and inter-site Coulomb interactions, respectively.

The low-energy $2a_1$ molecular-orbital model was constructed using Wannier90\cite{W90}. Screened Coulomb interactions $U$ and $V$ were evaluated using the constrained random-phase approximation (cRPA)\cite{cRPA} implemented in VASP, with 288 bands included in the GW calculation to ensure convergence. In cRPA, the screening channels internal to the target Wannier subspace are excluded from the polarization, yielding the effective interaction appropriate for the low-energy model. Although the screened interaction is frequency dependent in principle, it is nearly constant over the low-energy window relevant here and is therefore represented by static values, following Ref.~\cite{Calc2023,Calc2025}. The hopping and interaction parameters are listed in Table~\ref{tab:combined_parameters}.

\begin{table}[!htbp]
    \centering
    \caption{Hopping amplitudes ($t$) and screened Coulomb interactions ($U, V$) of the low-energy $2a_1$ model for $\mathrm{Nb}_3\mathrm{Cl}_8$, in meV. Here, $t_2^{\mathrm{(intra)}}$ and $V_2^{\mathrm{(intra)}}$ represent the nearest-neighbor intralayer hopping and interaction, respectively. The on-site Coulomb repulsion is denoted by $U$. The parameters $t_0^{\mathrm{(inter)}}, V_0^{\mathrm{(inter)}}$ and $t_1^{\mathrm{(inter)}}, V_1^{\mathrm{(inter)}}$ correspond to the two inequivalent interlayer connections: $(0,0,0;2)$ to $(0,0,1;1)$ and $(0,0,0;1)$ to $(0,0,0;2)$, respectively, following the two-site unit-cell convention introduced below.}
    \label{tab:combined_parameters}
    \begin{tabular}{cccccccc}
        \hline\hline
        phase & $t_0^{\,\mathrm{(inter)}}$ & $t_1^{\,\mathrm{(inter)}}$ & $t_2^{\,\mathrm{(intra)}}$ & $U$ & $V_0^{\,\mathrm{(inter)}}$ & $V_1^{\,\mathrm{(inter)}}$ & $V_2^{\,\mathrm{(intra)}}$ \\
        \hline
        $\alpha$ & -17.3  & -16.3 & 22.1 & 1394.8 & 329.5 & 287.5 & 467.4 \\
        $\beta$  & -136.7 & -16.4 & 18.8 & 1445.5 & 382.8 & 305.3 & 485.2 \\
        \hline\hline
    \end{tabular}
\end{table}

To avoid double counting in Eq.~\eqref{eq:free_energy_total}, the $2a_1$ low-energy contributions already captured by the DFT reference must be systematically subtracted. We implement a Hartree-level double-counting (DC) scheme consistent with the nonmagnetic DFT ground state, decomposing the correction into one-body kinetic, on-site, and inter-site Coulomb terms. For the one-body terms, the DC contribution is defined strictly as the nonlocal hopping energy of the $2a_1$ Wannier subspace, omitting the on-site component:
\begin{equation}
    \label{eq:E_hop_DC}
    E_{\mathrm{hop}}^{\mathrm{DC}}=\frac{2}{N_k}\sum_{\mathbf{k},\nu}^{\mathrm{occ}}\left\langle \psi_{\mathbf{k}\nu}\middle| \hat h_W(\mathbf{k})-\hat h_{\mathrm{on}}\middle| \psi_{\mathbf{k}\nu}\right\rangle .
\end{equation}
Here $\hat h_W(\mathbf{k})$ is the Wannier Hamiltonian, $\hat h_{\mathrm{on}}$ its on-site component, and $\psi_{\mathbf{k}\nu}$ the occupied eigenstates. Subtracting this quantity isolates the suppressed kinetic scale of the correlated Mott state by counteracting the artificial intercluster delocalization inherent to the Kohn-Sham reference.

For the on-site Coulomb interaction, the nonmagnetic DFT reference has one half-filled $2a_1$ orbital on each cluster, with $\langle \hat n_{i\uparrow}\rangle=\langle \hat n_{i\downarrow}\rangle=1/2$ and $\langle \hat n_i\rangle=1$. With the correlated subspace comprising a single equivalent orbital per site, orbital polarization is strictly absent. Consequently, the Hartree-level DC energy identically matches the Around-Mean-Field (AMF) functional:
\begin{equation}
    E_U^{\mathrm{DC}}=\sum_i U_i \langle \hat n_{i\uparrow}\rangle \langle \hat n_{i\downarrow}\rangle=\sum_i \frac{U_i}{4}.
\end{equation}
For the inter-site Coulomb interaction, the Hartree contribution at half filling is $\sum_{\langle i,j\rangle}V_{ij}$. We absorb this constant into the model Hamiltonian by replacing $V_{ij}\hat n_i\hat n_j$ with $V_{ij}(\hat n_i\hat n_j-1)$, rather than introducing a separate term. The explicit DC correction is thus restricted to
\begin{equation}
    \label{eq:E_DC}
    E_{\mathrm{DC}}=E_{\mathrm{hop}}^{\mathrm{DC}}+E_U^{\mathrm{DC}},
\end{equation}
and the shifted Hubbard Hamiltonian becomes
\begin{equation}
    \label{eq:H_Hub2}
    \begin{aligned}
    \hat H_{\mathrm{Hub}}={}&
    \sum_{\langle i,j\rangle,\sigma}
    \left(t_{ij}\hat c_{i\sigma}^{\dagger}\hat c_{j\sigma}+\mathrm{H.c.}\right)
    +\sum_i U_i\hat n_{i\uparrow}\hat n_{i\downarrow} \\
    &+\sum_{\langle i,j\rangle}V_{ij}(\hat n_i\hat n_j-1).
    \end{aligned}
\end{equation}

\subsection{Effective Heisenberg model}

Because the transition occurs at a temperature scale much lower than the charge-excitation energy of the half-filled $2a_1$ manifold, we further reduce the extended Hubbard model to an effective low-energy Heisenberg model:
\begin{equation}
    \label{eq:H_Heis}
    \hat{H}_\text{Hei} = \sum_{\langle i, j \rangle} J_{ij}\left(\mathbf{S}_i \cdot \mathbf{S}_j - \frac{1}{4}\right).
\end{equation}
The exchange parameter $J = 4t^{2}/(U-V)$ corresponds to the antiferromagnetic superexchange interaction between neighboring sites (see Supplementary Section 1 for details). The resulting exchange parameters are summarized in Table~\ref{tab:exchange_parameters}.

\begin{table}[!htbp]
    \centering
    \caption{Effective exchange parameters (meV) in $\mathrm{Nb}_3\mathrm{Cl}_8$, derived from the extended Hubbard model. The nearest-neighbor intralayer exchange $J_2^{\mathrm{intra}}$ and the two inequivalent interlayer exchanges $J_0^{\mathrm{inter}}$ and $J_1^{\mathrm{inter}}$ are evaluated as $J = 4t^2/(U-V)$.}
    \label{tab:exchange_parameters}
    \begin{tabular}{cccc}
        \hline\hline
        phase & $J_0^{\,\mathrm{(inter)}}$ & $J_1^{\,\mathrm{(inter)}}$ & $J_2^{\,\mathrm{(intra)}}$ \\
        \hline
        $\alpha$ & 1.12 & 0.96 & 2.12 \\
        $\beta$ & 70.34 & 0.94 & 1.58 \\
        \hline\hline
    \end{tabular}
\end{table}

For the $\alpha$ phase, we solved the $S = 1/2$ Heisenberg model using the linked-cluster high-temperature series expansion (HTSE) implemented in the \texttt{HTSE-code} package\cite{HTSE-code}. The series was computed to tenth order by summing all connected graphs on the lattice, which yields the cumulants entering the expansion of $\ln Z$. From these cumulants we obtained the free energy, entropy, and internal energy, and then extrapolated the free-energy and entropy series using the $P[4,4]$ and $P[4,6]$ Pad\'e approximants, respectively (see Supplementary Section 2 for details). Note that while HTSE provides reliable results at elevated temperatures, it cannot capture a potential QSL state that might emerge at very low temperatures. For the $\beta$ phase, we used a second-order perturbative expansion around the dimer-singlet limit. Because this expansion converges poorly at low temperature, the final low-temperature thermodynamics was taken from a simplified isolated-dimer model, as described in Supplementary Section 3.

\subsection{Enthalpy under uniaxial stress}

To describe the thermodynamics under applied uniaxial stress, and establish a baseline estimate for phase stability, we construct a generalized enthalpy at $T=0$ by adding the stress-work term $W_{\boldsymbol{\sigma}}$ to the Helmholtz free energy evaluated for structures obtained under imposed c-axis strain:
\begin{equation}
    \begin{aligned}
      \label{eq:enthalpy}
      H_\mathrm{tot}(T=0)=F_\mathrm{tot}(T=0)-V_0\sum_{i,j}\sigma_{ij}\epsilon_{ij} 
    \end{aligned}
    \end{equation}
We performed additional VASP calculations under uniaxial c-axis strain and extracted the corresponding pressure from the relaxed stress tensor\cite{elasticity}. The stress-work term is written as
\begin{equation}
    \label{eq:pressure}
    -V_{0}\sum_{ij}\sigma_{ij}\epsilon_{ij}=V_{0}P_{z}\epsilon_{zz}=-P_{z}A^{(xy)}_{0}(z_{0}-z).
\end{equation}
The generalized enthalpy at $T=0$ was then constructed from the DFT ground-state energy, the phonon zero-point energy, the zero-temperature internal energy of the Hubbard model, and the stress-work contribution. Pressure-induced renormalization of the finite-temperature phonon and spin entropies is not included in the present estimate of the critical pressure.

\section{Results and Discussion}

\subsection{Crystal structure and phase transition}

Experimentally, the high-temperature $\alpha$ phase of $\mathrm{Nb}_3\mathrm{Cl}_8$ crystallizes in $\mathrm{P\bar{3}m1}$ (No. 164). The low-temperature $\beta$ phase has been assigned to several closely related structures, including $\mathrm{C2/m}$ (No. 12)\cite{Rearrangement}, $\mathrm{R3}$ (No. 146)\cite{Disproportionation}, and $\mathrm{R\bar{3}m}$ (No. 166)\cite{suppression}. These structures differ in symmetry and internal distortions but share the same essential stacking motif with interlayer rearrangement. Among the reported $\beta$-phase candidates, $\mathrm{R\bar{3}m}$ has the highest symmetry. The lower symmetry structures introduce additional subtle distortions: the $\mathrm{C2/m}$ structure contains unequal intracluster $\mathrm{Nb}-\mathrm{Nb}$ bonds that break the threefold rotational symmetry, whereas the $\mathrm{R3}$ structure makes adjacent layers inequivalent and therefore breaks inversion symmetry\cite{Calc2023,NoncentroMottPRB}. Our DFT relaxations show that these additional distortions have only a minor energetic effect: the relaxed $\mathrm{R3}$ and $\mathrm{C2/m}$ structures are lower in energy than $\mathrm{R\bar{3}m}$ by 4 meV and 2 meV per unit cell, respectively. Accordingly, we use $\mathrm{R\bar{3}m}$ as the representative $\beta$-phase structure in the following calculations.

\begin{figure*}[t]
    \centering
    \includegraphics[width=0.8\textwidth]{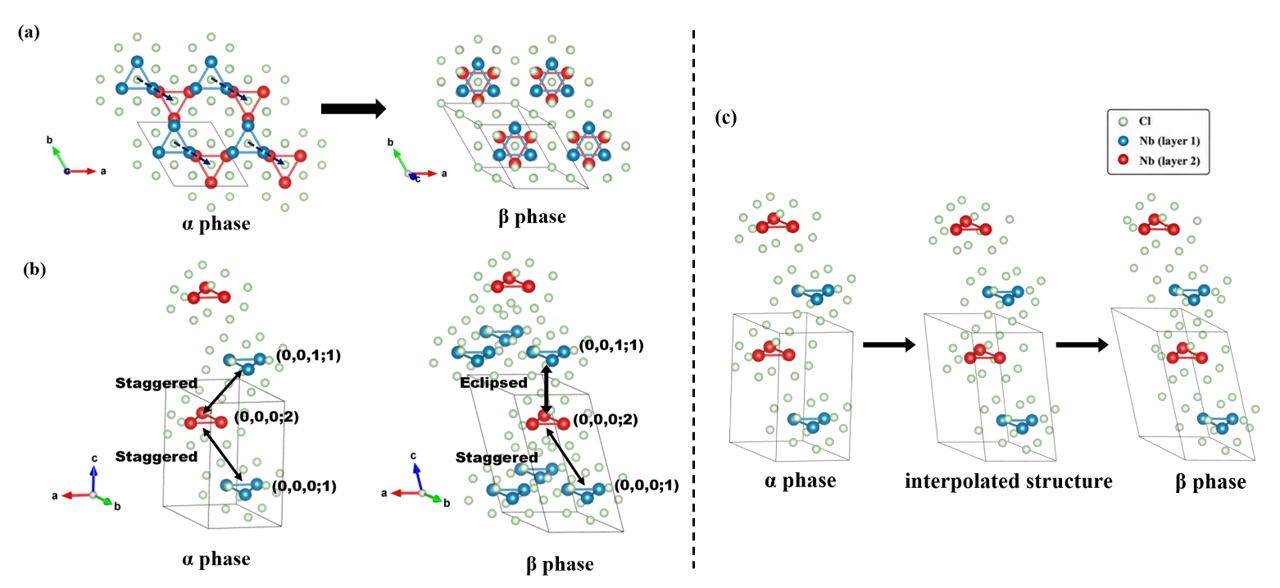}
    \caption{Crystal structures of $\mathrm{Nb}_3\mathrm{Cl}_8$ in the $\alpha$ and $\beta$ phases and the interpolation path between them. Light green spheres denote Cl atoms, while blue and red spheres denote Nb atoms in adjacent layers. (a) Top view of the interlayer slip from the staggered $\alpha$ stacking to the partially eclipsed $\beta$ stacking. The black arrow indicates the dominant structural displacement across the transition. (b) Side view of the corresponding stacking sequences. Here $(R,\mathrm{site})$ labels the lattice translation $R$ and the cluster site within the unit cell. The $\alpha$ phase has uniform staggered $AA^\prime$ stacking, whereas the $\beta$ phase adopts an $AA^\prime BB^\prime CC^\prime$ stacking sequence\cite{Tunable}, alternating between staggered and eclipsed interlayer pairs. (c) Representative structures along the interpolation path. The intermediate structures are generated by linearly interpolating both lattice parameters and internal atomic coordinates between the $\alpha$ and $\beta$ phase structures.}
    \label{fig:slip_stack_interp}
\end{figure*}

The structural phase transition from the $\alpha$ phase to the $\beta$ phase is dominated by an interlayer slip, as shown from the top view in Fig.~\ref{fig:slip_stack_interp}(a). For a direct comparison with the $\alpha$ phase, we describe the $\beta$-phase structure using a monoclinic cell analogous to the $\mathrm{C2/m}$ setting, rather than its primitive rhombohedral cell, so that the two phases can be compared within the same cell framework. Although the transition is accompanied by changes in the lattice parameters and internal atomic coordinates, including modest changes in the Nb-Cl-Nb bond angles\cite{Rearrangement}, the dominant structural change is an in-plane shift between neighboring layers. This slip changes the stacking geometry along the c axis from the uniform $\text{AA}'$ stacking of the $\alpha$ phase to the $\text{AA}'\text{BB}'\text{CC}'$ sequence of the $\beta$ phase [Fig.~\ref{fig:slip_stack_interp}(b)]\cite{Tunable}. In the $\alpha$ phase, adjacent cluster pairs are uniformly staggered, leading to relatively weak and nearly uniform interlayer coupling. In the $\beta$ phase, the stacking alternates between staggered and eclipsed interlayer pairs, producing alternating weak and strong interlayer couplings and promoting c-axis dimerization between neighboring clusters. To connect these two endpoint structures, we construct a representative interpolation path by linearly interpolating both lattice parameters and internal atomic coordinates between the relaxed $\alpha$- and $\beta$-phase structures [Fig.~\ref{fig:slip_stack_interp}(c)].

\begin{figure}[!htbp]
    \centering
    \includegraphics[width=0.8\columnwidth]{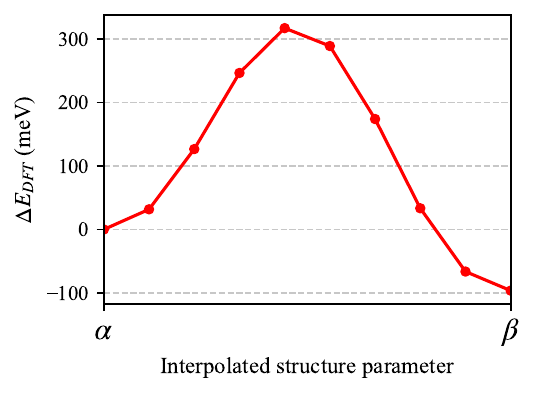}
    \caption{DFT total-energy difference per unit cell, $\Delta E_\mathrm{DFT}=E_\mathrm{DFT}-E_\mathrm{DFT}^{\alpha}$, versus the interpolated structural parameter between the $\alpha$ and $\beta$ phases. The $\alpha$-phase energy is taken as the reference.}
    \label{fig:DFT_energy}
\end{figure}

Along this interpolation path, the DFT total energy exhibits a sizable barrier of about 300 meV per two-site unit cell [Fig.~\ref{fig:DFT_energy}]. Such a substantial energy barrier is consistent with the hysteresis effects observed in experiments\cite{Rearrangement}. However, this value serves merely as an estimate along the chosen path and should not be regarded as the exact kinetic barrier of the actual phase transition. 

\subsection{Thermodynamic driving forces}

At zero external stress, the thermodynamic driving forces for the phase transition are analyzed by decomposing the free energy into four contributions: the DFT total energy without spin polarization, the double counting correction, the phonon free energy, and the Hubbard model free energy of the correlated $2a_1$ manifold. These contributions are discussed separately below.

At the non-spin-polarized DFT level, the $\beta$ phase is lower than the $\alpha$ phase by about 100 meV per unit cell [Fig.~\ref{fig:DFT_energy}], reflecting the enhanced bonding-antibonding splitting produced by the alternating interlayer coupling in the $\beta$ structure. This value, however, should not be interpreted as the correlated $\alpha$-$\beta$ energy splitting, because standard DFT cannot properly describe the half-filled $2a_1$ Mott subspace (see Supplementary Section 1 for details). In our free-energy construction, this low-energy contribution is treated explicitly by the effective Hubbard model, while the DFT energy enters only as the remaining structural background after the Hartree-level double-counting correction in Eq.~\eqref{eq:E_DC}. Although bond-order-dependent Fock terms are not included in this Hartree-level double-counting scheme, they are expected to be much smaller than the Hartree term.

\begin{figure*}[t]
    \centering
    \includegraphics[width=0.8\textwidth]{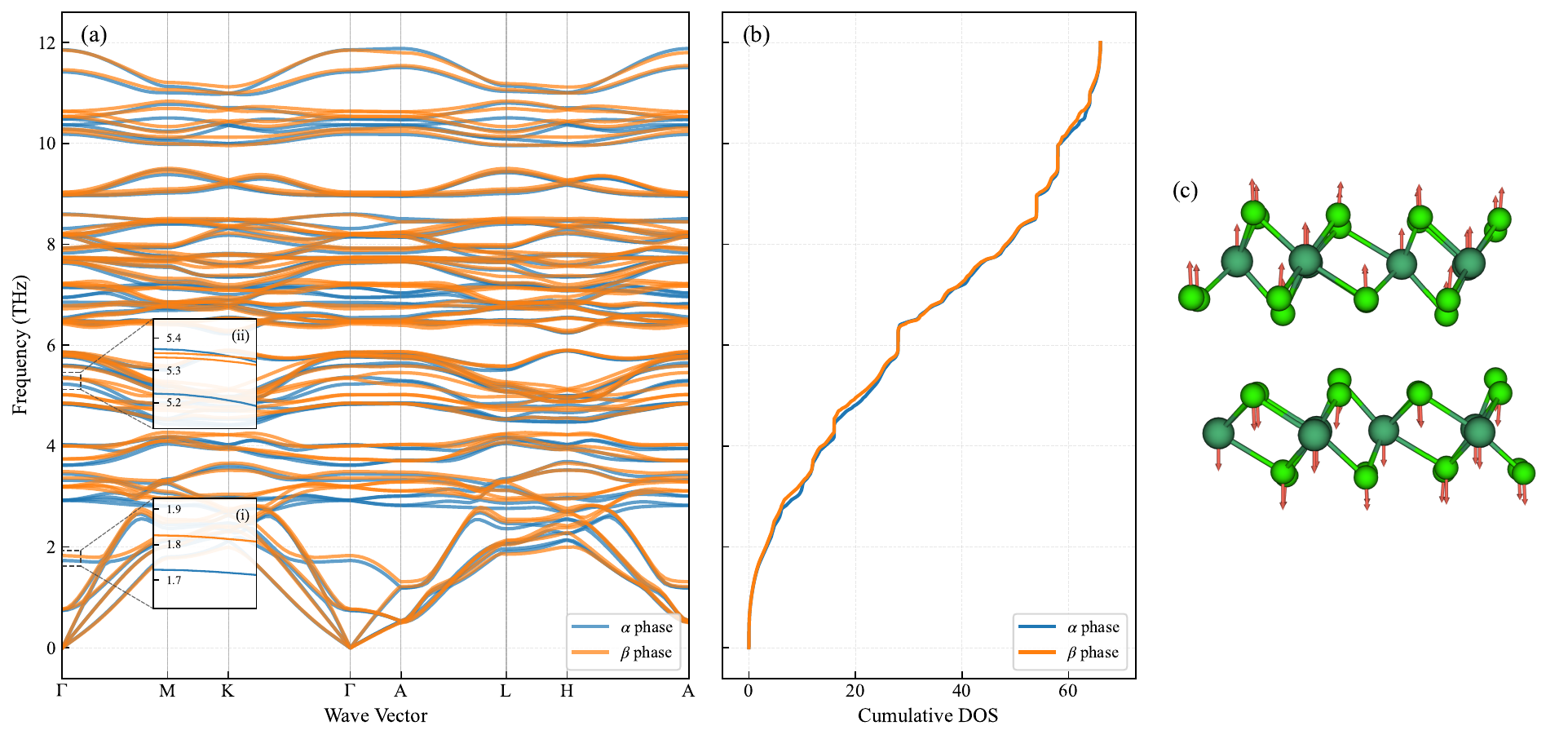}
    \caption{(a) Phonon dispersions of the $\alpha$ phase (blue) and $\beta$ phase (orange) of $\mathrm{Nb}_3\mathrm{Cl}_8$, plotted along the high-symmetry path of the $\alpha$ phase. Inset (i) highlights the $\Gamma$-point Layer-Breathing Mode, with frequencies of 1.729 THz in the $\alpha$ phase and 1.826 THz in the $\beta$ phase. Inset (ii) highlights the $\Gamma$-point modes near 5.3 THz for comparison with experimentally observed phonon modes. In the $\alpha$ phase, the $A_{1g}$ and $A_{2u}$ modes are at 5.228 and 5.366 THz, respectively; in the $\beta$ phase, the $A_{2u}$ and $A_{1g}$ modes are at 5.341 and 5.354 THz, respectively. (b) Corresponding cumulative phonon density of states. The $\alpha$-phase curve is shifted slightly toward lower frequencies, indicating an overall phonon softening relative to the $\beta$ phase. (c) Illustration of the Layer-Breathing Mode highlighted in inset (i) of panel (a); light green and dark green spheres denote Cl and Nb atoms, respectively.}
    \label{fig:Phonon_dos}
\end{figure*}

\begin{figure}[!htbp]
    \centering
    \includegraphics[width=0.8\columnwidth]{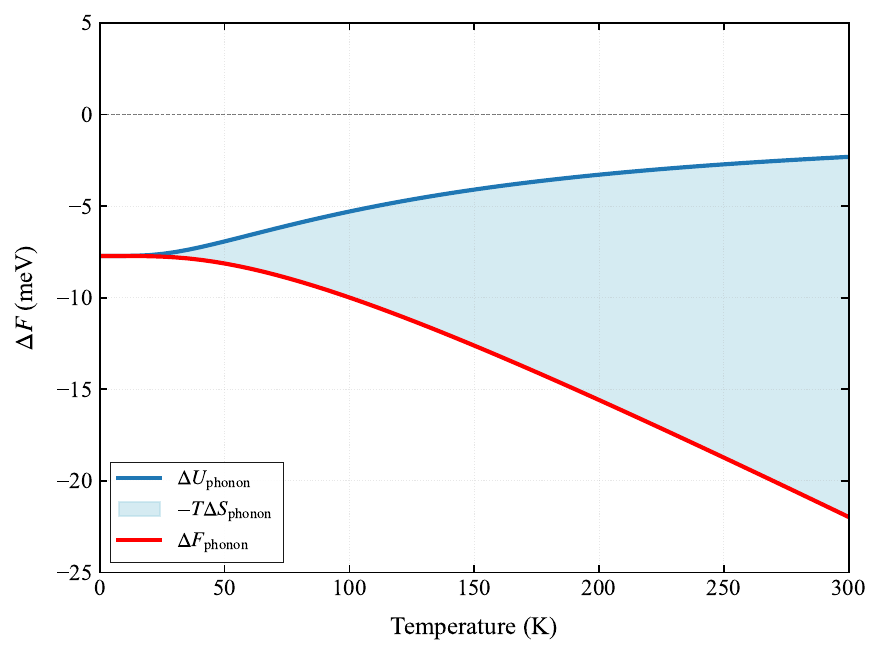}
    \caption{Phonon Helmholtz free-energy difference per unit cell, $\Delta F_{\mathrm{phonon}} = F_{\mathrm{phonon}}^\alpha - F_{\mathrm{phonon}}^\beta$, versus temperature. The decomposition includes the phonon internal-energy difference $\Delta U_{\mathrm{phonon}}$ (blue line) and the entropy contribution $-T\Delta S_{\mathrm{phonon}}$ (blue area). At $T=0$, $\Delta U_{\mathrm{phonon}}$ reduces to the phonon zero-point-energy difference $\Delta E_{\mathrm{zpe}} = E_{\mathrm{zpe}}^\alpha - E_{\mathrm{zpe}}^\beta$.}
    \label{fig:Phonon_free_energy}
\end{figure}

Beyond this electronic background, lattice vibrations provide a temperature-dependent contribution to the relative stability of the two phases. The phonon spectra and cumulative phonon DOS calculated from the relaxed structures are shown in Fig.~\ref{fig:Phonon_dos}, and the corresponding phonon free-energy difference is presented in Fig.~\ref{fig:Phonon_free_energy}.

A comparison of the phonon DOS between the $\alpha$ and $\beta$ phases [Fig.~\ref{fig:Phonon_dos}(b)] reveals a subtle overall phonon softening in the $\alpha$ phase. This trend originates from their structural distinctions and is exemplified by the layer-breathing mode shown in inset (i) of Fig.~\ref{fig:Phonon_dos}(a) and Fig.~\ref{fig:Phonon_dos}(c). Characterized by slightly smaller lattice volume and enhanced interlayer coupling, the $\beta$ phase exhibits a larger effective restoring force constant $k$, which shifts the phonon frequencies upward according to $\omega \sim \sqrt{k/M}$. In contrast, the weaker interlayer coupling in the $\alpha$ phase yields reduced restoring forces, thereby inducing softer phonon modes. Such behavior agrees with the overall $\beta$-phase phonon hardening reported in Refs.~\cite{RamanNb3Cl8,THzNb3X8}. The $\Gamma$-point Raman-active mode near 5.3 THz, shown in inset (ii) of Fig.~\ref{fig:Phonon_dos}(a), also provides a direct comparison with experimental observations: the $A_{1g}$ mode is located at 5.228 THz (174.4 cm$^{-1}$) in the $\alpha$ phase and shifts to 5.354 THz (178.6 cm$^{-1}$) in the $\beta$ phase. This frequency increase is consistent with the variations reported in Ref.~\cite{suppression}. 

This structural softening lowers the zero-point energy and increases the entropy of the $\alpha$ phase, promoting its high-temperature stability, analogous to the entropy-driven stabilization discussed for $\mathrm{ZrO_{2}}$\cite{ZrO2} and $\mathrm{VO_{2}}$\cite{VO2,VO2_2}. At finite temperatures, while more phonon modes are populated in the $\alpha$ phase, their reduced average energy creates a complex temperature dependence for the phonon internal energy. Nevertheless, the entropic contribution remains dominant, driving a more rapid decrease in the phonon free energy of the $\alpha$ phase [Fig.~\ref{fig:Phonon_free_energy}].

\begin{figure}[!htbp]
    \centering
    \includegraphics[width=0.8\columnwidth]{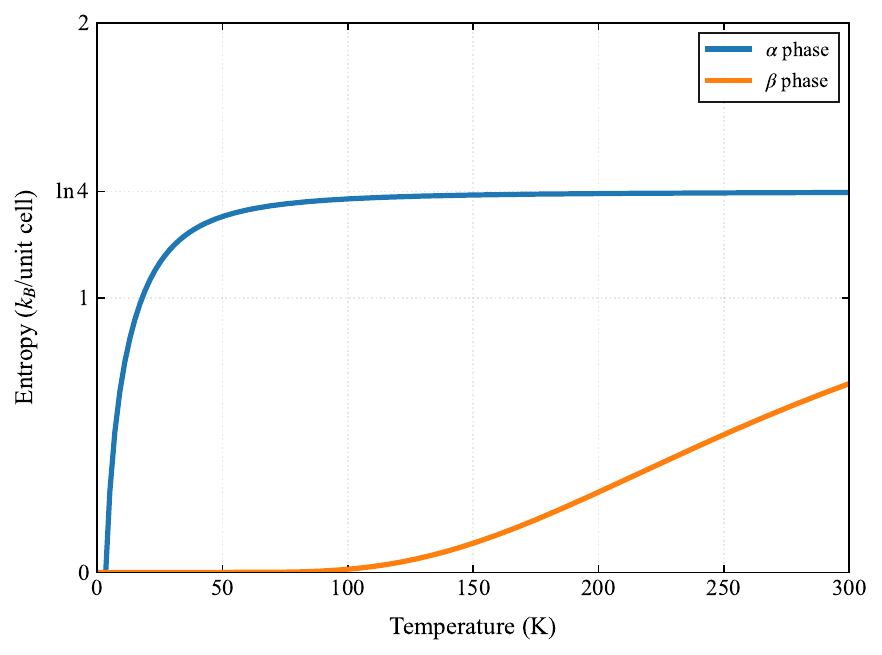}
    \caption{Spin entropy per unit cell of the effective Heisenberg model versus temperature. For the $\alpha$ phase (blue), the entropy is obtained from the HTSE using the $P[4,6]$ Pad\'e approximant. For the $\beta$ phase (orange), the entropy is obtained from the isolated-dimer model.}
    \label{fig:Hubbard_entropy}
\end{figure}

\begin{figure}[!htbp]
    \centering
    \includegraphics[width=0.8\columnwidth]{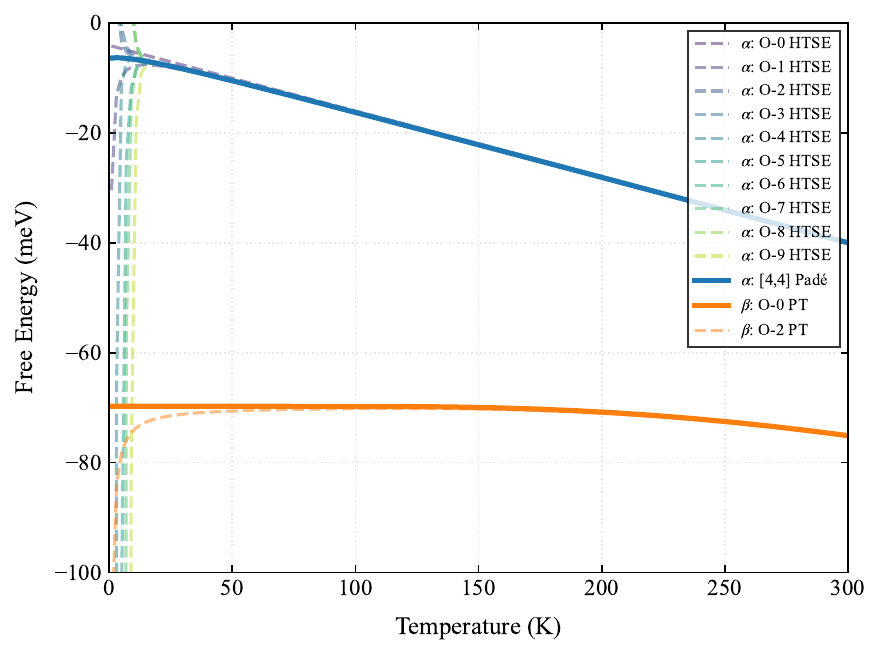}
    \caption{Spin Helmholtz free energy per unit cell of the effective Heisenberg model versus temperature. For the $\alpha$ phase, the $P[4,4]$ Pad\'e approximant (blue solid line) is shown together with the raw HTSE truncated at different orders (colored dashed lines). For the $\beta$ phase, the isolated-dimer result (orange solid line) is compared with the second-order perturbative expansion around the dimer limit (orange dashed line).}
    \label{fig:Hubbard_free_energy}
\end{figure}

The spin thermodynamics provides the second major entropy contribution. In the $\beta$ phase, the strong interlayer hopping $t_0^{\mathrm{(inter)}}=-136.7$ meV produces a large exchange $J_0^{\mathrm{(inter)}}=70.34$ meV (Tables~\ref{tab:combined_parameters} and \ref{tab:exchange_parameters}), driving interlayer singlet formation and quenching the local moment. In contrast, the interlayer couplings in the $\alpha$ phase are relatively weak and comparable to the intralayer scale, so the system retains paramagnetic spin entropy near the transition. The resulting entropy and free energy of the effective Heisenberg model are shown in Figs.~\ref{fig:Hubbard_entropy} and \ref{fig:Hubbard_free_energy}.

Combining the DFT structural background, DC correction, harmonic phonon free energy, and the low-energy Hubbard/Heisenberg contribution, we obtain the total Helmholtz free energy at zero external stress according to Eq.~\eqref{eq:free_energy_total}. Within the present approximations, the free-energy crossing occurs at $T_c \approx 113\,\mathrm{K}$, on the same $\sim 100\,\mathrm{K}$ scale as the experimental transition temperature \cite{Rearrangement,observation,Disproportionation}.

\begin{figure}[!htbp]
    \centering
    \includegraphics[width=0.8\columnwidth]{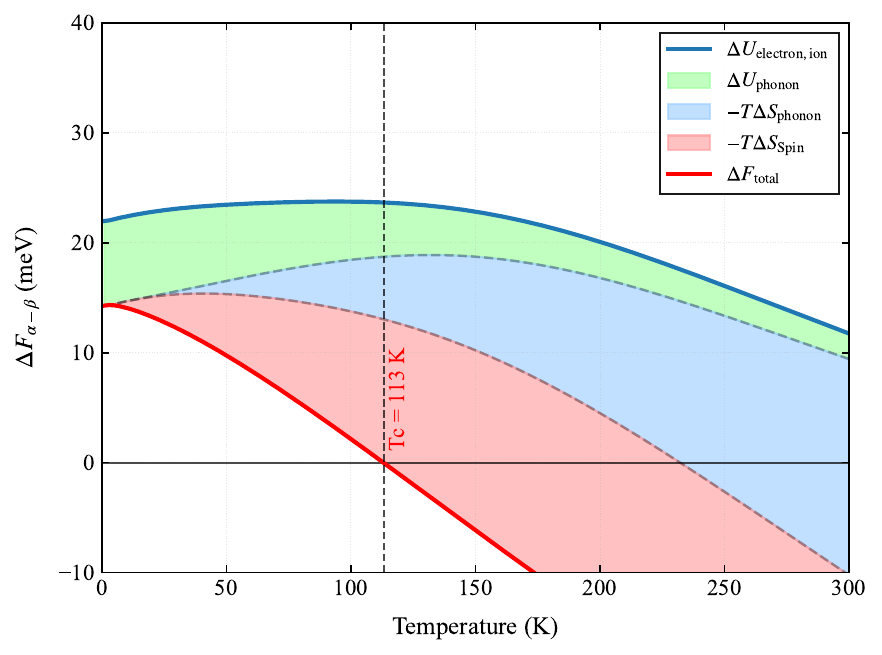}
    \caption{Total Helmholtz free energy difference per unit cell between the $\alpha$ and $\beta$ phases, $\Delta F_{\mathrm{total}} = F_{\mathrm{total}}^\alpha - F_{\mathrm{total}}^\beta$, as a function of temperature at zero pressure. The total difference is decomposed into the electronic and ionic energy contribution $\Delta E_{\mathrm{elec+ion}}=\Delta(E_\mathrm{DFT}-E_\mathrm{DC}+E_\mathrm{Hub})$, the phonon internal energy contribution $\Delta E_{\mathrm{phonon}}$, the phonon entropy contribution $-T\Delta S_{\mathrm{phonon}}$, and the spin entropy contribution $-T\Delta S_{\mathrm{spin}}$, as indicated by the lines and shaded regions. The condition $\Delta F_{\mathrm{total}}=0$ yields an estimated transition temperature of $T_c \approx 113\,\mathrm{K}$ within the present approximations.}
    \label{fig:Total_free_energy}
\end{figure}

The resulting balance shows that the high-temperature stability of the $\alpha$ phase is mainly provided by phonon and spin entropies [Fig.~\ref{fig:Total_free_energy}], whereas the low-temperature $\beta$ phase is favored by the internal-energy gain associated with interlayer dimerization. In the $\beta$ phase, the alternating interlayer couplings generate strongly bonded cluster dimers, which favor spin-singlet formation and suppress the spin entropy, while the stronger interlayer coupling also hardens the phonons and reduces the phonon entropy contribution.

Experimentally, the entropy change across the transition, as inferred from specific-heat measurements, is approximately 85\% of $S = k_B \ln 4$ per unit cell\cite{Rearrangement}. This is the same entropy scale as the two-site unit cell used here, whose high-temperature limit is $2k_B\ln 2 = k_B\ln 4$. When possible impurity spins and remnant $\alpha$-phase contributions in the $\beta$ phase are taken into account\cite{Rearrangement,pressure}, this value is consistent with our calculations. In addition, recent HREELS measurements found a quasi-2D linearly dispersing exciton in the $\alpha$ phase and split quasi-3D excitonic bands in the $\beta$ phase\cite{ExcitonPRL2026}, directly supporting the picture that the transition is governed by a substantial reconstruction of interlayer coupling rather than by a weak perturbation of an otherwise unchanged electronic structure.

Several approximations affect the quantitative transition temperature. First, the finite-temperature vibrational free energy is evaluated within the harmonic approximation using phonon spectra calculated for the zero-temperature structures. Temperature-induced phonon renormalization and possible anharmonic effects of low-frequency interlayer modes are therefore not included explicitly. Second, these phonon spectra are obtained from DFT reference states, in which strong correlations in the low-energy $2a_1$ subspace are not properly described, and electron-phonon coupling beyond static structural relaxation is neglected. Third, the HTSE provides a finite-temperature description near the transition regime, but its accuracy at lower temperatures relies on the Pad\'e continuation; this mainly affects the region below about 20 K in Fig.~\ref{fig:Hubbard_free_energy}—where the method fails to resolve a possible QSL state—while the influence near $T_c$ is much smaller. Finally, no separate entropy contribution from itinerant electronic excitations is included, but this omission is expected to be minor because both phases are insulating. More subtle material-specific effects, including lattice distortions, disorder, anharmonic low-frequency modes, and double-counting ambiguities in the correlated total energy, are also beyond the present treatment.

\subsection{Effect of uniaxial pressure}

Uniaxial pressure along the c-axis serves as an experimental approach to suppressing the structural transition and may stabilize the QSL-candidate $\alpha$ phase down to low temperatures\cite{pressure}. By evaluating the $T=0$ enthalpy difference $\Delta H = H_\alpha - H_\beta$ from Eq.~\eqref{eq:enthalpy}, we estimate a semiquantitative threshold for the pressure-induced suppression of the $\beta$ phase.

\begin{figure}[t]
    \centering
    \includegraphics[width=0.8\columnwidth]{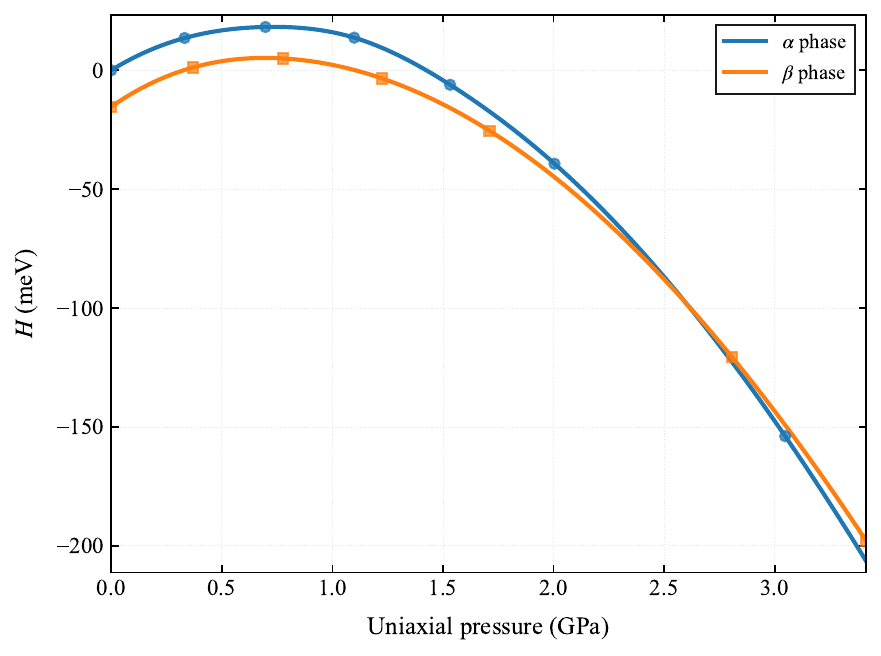}
    \caption{Generalized enthalpy per unit cell, $H$, versus c-axis uniaxial pressure at $T=0$. At low pressure, the increase in internal energy dominates for both phases, whereas at higher pressure the stress-work term $W_{\boldsymbol{\sigma}}$ drives the downward trend. The crossing marks stabilization of the $\alpha$ phase. The $\alpha$-phase enthalpy at $P=0$ is used as the reference.}
    \label{fig:Pressure_free_energy}
\end{figure}

The pressure effects can be understood from the distinct structural responses of the two phases to uniaxial compression. Under c-axis pressure, the generalized enthalpy reflects a competition between the increasing internal energy and the decreasing stress-work term $W_{\boldsymbol{\sigma}}$\cite{elasticity}, with the latter becoming dominant at higher pressures [Fig.~\ref{fig:Pressure_free_energy}]. Owing to its stronger interlayer coupling, the $\beta$ phase is less compressible than the $\alpha$ phase. As a result, the generalized enthalpy of the $\alpha$ phase is reduced more strongly under the same c-axis pressure, thereby favoring its stabilization.

\begin{figure}[t]
    \centering
    \includegraphics[width=0.8\columnwidth]{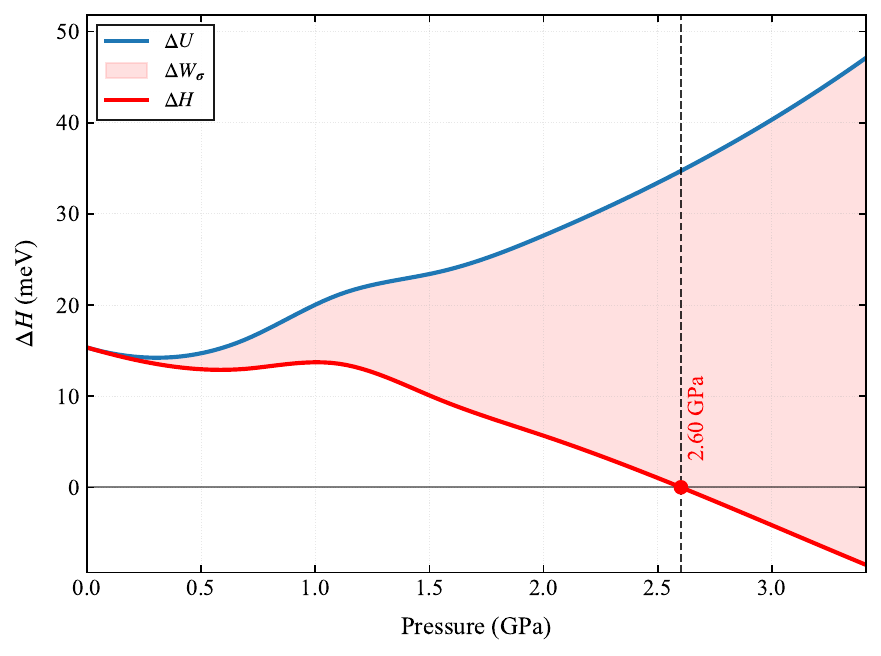}
    \caption{Decomposition of the generalized enthalpy difference per unit cell between the $\alpha$ and $\beta$ phases. The total difference $\Delta H = H_\alpha - H_\beta$ (red line) is resolved into the internal-energy part $\Delta U$ (blue line) and the stress-work contribution $\Delta W_{\boldsymbol{\sigma}}$ (red area). The zero crossing yields an estimated critical pressure of $P_c \approx 2.60$ GPa within the present $T=0$ treatment.}
    \label{fig:Pressure_delta_free_energy}
\end{figure}

Our $T=0$ generalized-enthalpy calculations [Fig.~\ref{fig:Pressure_delta_free_energy}] indicate that the structural transition is fully suppressed above a critical pressure of about 2.6 GPa. This value should be viewed as semiquantitative, since finite-temperature phonon and spin entropic contributions under pressure are not included. At 1.8 GPa, comparable to the experimental range, the transition is only partially suppressed, consistent with magnetic-susceptibility measurements\cite{pressure}.   

These results indicate that uniaxial compression along the c axis raises the enthalpy of the $\beta$ phase relative to the $\alpha$ phase, thereby suppressing the transition and stabilizing the $\alpha$ structure at low temperatures. This may provide a route to observing the candidate QSL behavior of the $\alpha$ phase. Conversely, tensile strain along the c axis, which can be qualitatively related to in-plane compression, may stabilize the $\beta$ phase and increase the transition temperature, making its crystal and electronic structures accessible over a broader experimental temperature range.

\section{Conclusion}

In this study, we investigated the microscopic mechanism of the first-order structural phase transition in $\mathrm{Nb_3Cl_8}$ by combining first principles calculations with an extended Hubbard model. We established a free energy framework incorporating lattice, electronic, and spin degrees of freedom and showed that the transition is jointly driven by phonon and spin entropy. The high-temperature $\alpha$ phase is stabilized by phonon softening and paramagnetic spin entropy, whereas the low-temperature $\beta$ phase loses entropy through phonon hardening and interlayer spin singlet formation. Within the present approximations, the free energy crossing occurs at $\sim 113$ K, on the same temperature scale as experiment, which provides semiquantitative support for this entropy-energy balance rather than a precise prediction.

We further found that c-axis uniaxial pressure suppresses the transition by producing a larger reduction in the generalized enthalpy of the $\alpha$ phase within the present $T=0$ treatment.

More broadly, our results offer a framework for understanding structural transitions in cluster Mott insulators, where lattice, electronic, and spin degrees of freedom are strongly intertwined.

\begin{acknowledgments}
    This work was supported by the National Natural Science Foundation of China (Grant Nos. 12188101, 12274436, and 11921004) and the National Key R\&D Program of China (Grant Nos. 2022YFA1403800 and 2023YFA1607400). H.W. acknowledges support from the New Cornerstone Science Foundation through the XPLORER PRIZE. Data management was performed using the MatElab platform of the Condensed Matter Physics Data Center, Chinese Academy of Sciences.
\end{acknowledgments}

\bibliographystyle{apsrev4-2}
\bibliography{ref}

\end{document}


\title{Supplemental Material for Entropy-Driven Structural Phase Transition in \texorpdfstring{Nb$_3$Cl$_8$}{Nb3Cl8} via DFT and an Effective Model}

\author{Chenjie Zhu}
\affiliation{Beijing National Laboratory for Condensed Matter Physics and Institute of Physics, Chinese Academy of Sciences, Beijing 100190, China}

\author{Shuai Zhang}
\affiliation{Institute of Theoretical Physics, Chinese Academy of Sciences, Beijing 100190, China}
\author{Zhong Fang}
\affiliation{Beijing National Laboratory for Condensed Matter Physics and Institute of Physics, Chinese Academy of Sciences, Beijing 100190, China}
\author{Zhijun Wang}
\affiliation{Beijing National Laboratory for Condensed Matter Physics and Institute of Physics, Chinese Academy of Sciences, Beijing 100190, China}
\author{Quansheng Wu}
\affiliation{Beijing National Laboratory for Condensed Matter Physics and Institute of Physics, Chinese Academy of Sciences, Beijing 100190, China}

\author{Hongming Weng}
\email{hmweng@iphy.ac.cn}
\affiliation{Beijing National Laboratory for Condensed Matter Physics and Institute of Physics, Chinese Academy of Sciences, Beijing 100190, China}

\date{\today}

\maketitle

\section{Derivation of the Effective Heisenberg Model}

We consider the extended Hubbard model
\begin{equation}
    \label{eq:Hubbard}
    \hat{H}_{\text{Hubbard}} = \underbrace{ \sum_{\langle i, j \rangle, \sigma} (t_{ij}\hat{c}_{i\sigma}^{\dagger} \hat{c}_{j\sigma} + \text{H.c.})}_{\text{Kinetic term}} + \underbrace{\sum_{i} U_i\hat{n}_{i\uparrow} \hat{n}_{i\downarrow}}_{\text{On-site } U} + \underbrace{\sum_{\langle i, j \rangle} V_{ij}(\hat{n}_i \hat{n}_j-1)}_{\text{Off-site } V},
\end{equation}
where \(t_{ij}\) is the hopping amplitude, \(U_i\) is the on-site Coulomb interaction, and \(V_{ij}\) is the nearest-neighbor intersite Coulomb interaction. At half filling and in the strong-coupling regime \(U-V \gg t,\, k_B T\), the low-energy physics is described by an effective Heisenberg model,
\begin{equation}
    \label{eq:Heisenberg}
    \hat{H}_{\text{Heisenberg}} = \sum_{\langle i, j \rangle} J_{ij}\left(\mathbf{S}_i \cdot \mathbf{S}_j - \frac{1}{4}\right).
\end{equation}

Because second-order perturbation theory involves processes spanning at most two sites, it is sufficient to analyze a minimal two-site problem. At half filling, the low-energy subspace is spanned by the four singly occupied states
\begin{equation}
    \{ \ket{1\uparrow,2\uparrow}, \ket{1\downarrow,2\downarrow}, \ket{1\uparrow,2\downarrow}, \ket{1\downarrow,2\uparrow} \}.
    \label{eq:low_states}
\end{equation}
The high-energy subspace consists of the two doubly occupied states
\begin{equation}
    \{ \ket{1\uparrow,1\downarrow}, \ket{2\uparrow,2\downarrow} \}.
    \label{eq:high_states}
\end{equation}

In the basis ordered as
\begin{equation}
    \{ \ket{1\uparrow,2\uparrow}, \ket{1\downarrow,2\downarrow}, \ket{1\uparrow,2\downarrow}, \ket{1\downarrow,2\uparrow}, \ket{1\uparrow,1\downarrow}, \ket{2\uparrow,2\downarrow} \},
    \label{eq:basis}
\end{equation}
the two-site extended Hubbard Hamiltonian reads
\begin{equation}
    H_{\text{Hubbard}} =
    \begin{pmatrix}
    0 & 0 & 0 & 0 & 0 & 0 \\
    0 & 0 & 0 & 0 & 0 & 0 \\
    0 & 0 & 0 & 0 & t & -t \\
    0 & 0 & 0 & 0 & -t & t \\
    0 & 0 & t & -t & U-V & 0 \\
    0 & 0 & -t & t & 0 & U-V
    \end{pmatrix}.
    \label{eq:Hubbard_matrix}
\end{equation}
The constant \(-1\) in the term \(V_{ij}(\hat{n}_i \hat{n}_j-1)\) of Eq.~\ref{eq:Hubbard} accounts for the constant intersite Hartree contribution discussed in the main text.

In the strong-coupling regime \(U-V \gg t\), the doubly occupied states act as virtual intermediate states in second-order hopping processes and generate an effective spin exchange \(J\). For the spin-parallel states \(\ket{1\uparrow,2\uparrow}\) and \(\ket{1\downarrow,2\downarrow}\), Pauli blocking forbids virtual hopping. The nontrivial processes therefore occur only in the subspace \(\{\ket{1\uparrow,2\downarrow}, \ket{1\downarrow,2\uparrow}\}\).

For the diagonal matrix element \(\bra{1\uparrow,2\downarrow} H_{\text{eff}} \ket{1\uparrow,2\downarrow}\), the two virtual processes
\(\ket{1\uparrow,2\downarrow} \rightarrow \ket{1\uparrow,1\downarrow} \rightarrow \ket{1\uparrow,2\downarrow}\) and
\(\ket{1\uparrow,2\downarrow} \rightarrow \ket{2\uparrow,2\downarrow} \rightarrow \ket{1\uparrow,2\downarrow}\)
each contribute \(t^2/[-(U-V)]\), giving a total of \(2t^2/[-(U-V)]\). The same result holds for \(\bra{1\downarrow,2\uparrow} H_{\text{eff}} \ket{1\downarrow,2\uparrow}\).

For the off-diagonal matrix element \(\bra{1\downarrow,2\uparrow} H_{\text{eff}} \ket{1\uparrow,2\downarrow}\), the two virtual processes
\(\ket{1\uparrow,2\downarrow} \rightarrow \ket{1\uparrow,1\downarrow} \rightarrow \ket{1\downarrow,2\uparrow}\) and
\(\ket{1\uparrow,2\downarrow} \rightarrow \ket{2\uparrow,2\downarrow} \rightarrow \ket{1\downarrow,2\uparrow}\)
each contribute \(t^2/(U-V)\), for a total of \(2t^2/(U-V)\).

In the basis \(\{\ket{1\uparrow,2\uparrow}, \ket{1\downarrow,2\downarrow}, \ket{1\uparrow,2\downarrow}, \ket{1\downarrow,2\uparrow}\}\), the Heisenberg Hamiltonian of Eq.~\ref{eq:Heisenberg} takes the form
\begin{equation}
    H_{\text{Heisenberg}} =
    \begin{pmatrix}
    0 & 0 & 0 & 0 \\
    0 & 0 & 0 & 0 \\
    0 & 0 & -\frac{J}{2} & \frac{J}{2} \\
    0 & 0 & \frac{J}{2} & -\frac{J}{2}
    \end{pmatrix}.
    \label{eq:Heisenberg_matrix}
\end{equation}
Matching the perturbative result to Eq.~\ref{eq:Heisenberg_matrix} gives the exchange coupling
\begin{equation}
    J = \frac{4t^2}{U-V}.
    \label{eq:J_formula}
\end{equation}
The same expression follows from Löwdin's degenerate perturbation theory\cite{lowdin},
\begin{equation}
    U_{mn}^{A} = H_{mn} + \sum_{\alpha \in B} \frac{H_{m\alpha}' H_{\alpha n}'}{E_0 - H_{\alpha\alpha}} + \ldots,
    \label{eq:lowdin}
\end{equation}
\begin{equation}
    H_{mn}' = H_{mn} (1 - \delta_{mn}),
\end{equation}
where $A$ denotes the degenerate low-energy subspace of singly occupied states, $B$ the high-energy subspace of doubly occupied states, $H$ the full Hamiltonian, and $U^{A}$ the effective Hamiltonian projected onto subspace $A$.

Diagonalizing the full two-site Hamiltonian provides an instructive cross-check for this minimal problem. The exact ground-state energy of the two-site Hamiltonian is
\begin{equation}
    E_g = -\frac{1}{2} \left( \sqrt{16 t^2 + (U-V)^2}- (U - V) \right).
    \label{eq:exact_energy}
\end{equation}
A corresponding unnormalized eigenvector is
\begin{equation}
    \left(0,\ 0,\ \frac{4 t}{\sqrt{16 t^2+(U-V)^2}-(U-V)},\ -\frac{4 t}{\sqrt{16 t^2+(U-V)^2}-(U-V)},\ -1,\ 1\right).
    \label{eq:exact_vector}
\end{equation}

\begin{itemize}
    \item In the strong-coupling limit \(U-V \gg t\): Expanding \(E_g\) [Eq.~\eqref{eq:exact_energy}] to leading order in \(t\) gives \(E_g \approx -4 t^2 / (U-V)\), and the ground state is proportional to \((0,0,1,-1,0,0)\). This is the two-site spin-singlet limit underlying the dimerized low-temperature state.
    \item In the weak-coupling limit \(U-V \ll t\): The ground-state energy approaches \(-2t\) for the sign convention used here, and the eigenvector is proportional to \((0,0,1,-1,-1,1)\), which is a bonding state with delocalized electron character.
\end{itemize}

This comparison illustrates the double-counting issue in standard density functional theory (DFT) for strongly correlated materials. In the large-\(U\) limit, the separate effects of hopping \(t\) and interaction \(U\) are reorganized into the superexchange scale \(J\) of Eq.~\ref{eq:J_formula}, so the correlated low-energy physics is much less sensitive to the bare one-particle parameters than the DFT total energy would suggest.

\section{High-Temperature Series Expansion of the Heisenberg Model}
Thermodynamic observables such as the specific heat and magnetic susceptibility follow from the Helmholtz free energy per spin,
\begin{equation}
    f = \frac{F}{N} = -\frac{k_B T}{N}\ln Z,
\end{equation}
where $Z =  \mathrm{Tr} \bigl(e^{-\beta \mathcal{H}}\bigr)$ is the partition function and $\beta = 1/(k_B T)$ is the inverse temperature. In the high-temperature regime, where $k_B T$ is large compared with the interaction strength, one can perform a systematic expansion in powers of $\beta$.

A direct expansion of the partition function gives
\begin{equation}
    Z =  \mathrm{Tr} \bigl(e^{-\beta \mathcal{H}} \bigr) = Z_0\sum_{m=0}^{\infty} \frac{(-\beta)^m}{m!} \langle \mathcal{H}^m \rangle,
\end{equation}
\begin{equation}
    Z_0= \mathrm{Tr} \bigl(\mathcal{I} \bigr).
\end{equation}

The moments $\langle \mathcal{H}^m \rangle = { \mathrm{Tr} \bigl(\mathcal{H}^m \bigr)}/{ \mathrm{Tr} \bigl(\mathcal{I}\bigr)}$ contain both connected and disconnected contributions. Expanding $Z$ directly in these moments produces terms that scale nonlinearly with the system size $N$. The standard approach therefore expands $\ln Z$, which removes disconnected diagrams order by order and yields the cumulant, or linked-cluster, expansion
\begin{equation}
    \begin{aligned}
    \frac{\ln Z}{N} &= \frac{1}{N} \ln \left[  \mathrm{Tr} \bigl(e^{-\beta \mathcal{H}} \bigr) \right] \\
    &= \frac{\ln Z_0}{N} + \frac{1}{N} \ln \left( 1 + \sum_{m=1}^{\infty} \frac{(-\beta)^m}{m!} \langle \mathcal{H}^m \rangle \right) \\
    &= \frac{\ln Z_0}{N} + \sum_{m=1}^{\infty} \frac{(-\beta)^m}{m!} \frac{\langle \mathcal{H}^m \rangle_c}{N}.
    \end{aligned}
\end{equation}

Here, $\langle \mathcal{H}^m \rangle_c$ denotes the $m$th-order cumulant. These cumulants are extensive ($\propto N$), which ensures a well-behaved thermodynamic-limit series. They are also invariant under constant shifts of the Hamiltonian, except for the first cumulant $\langle \mathcal{H} \rangle_c$.

We apply this construction to the isotropic Heisenberg model
\begin{equation}
    \mathcal{H} = \sum_{\langle i, j \rangle} J_{ij} \left( \mathbf{S}_i \cdot \mathbf{S}_j \right),
\end{equation}
where the sum runs over interacting bonds, $J_{ij}$ is the exchange coupling, and $\mathbf{S}_i$ are spin-$S$ operators. Evaluating the cumulants \(\langle \mathcal{H}^m \rangle_c\) then reduces to computing traces of products of spin operators over connected bond clusters.

Since the HTSE calculation uses the unshifted convention above, the absolute free energy compared with the shifted main-text Hamiltonian includes the constant correction $-\frac{1}{4}\sum_{\langle i,j\rangle}J_{ij}$, which does not affect the entropy or specific heat.

For spin \(1/2\), the constant term \(\ln Z_0/N\) is simply the high-temperature entropy of an unconstrained spin,
\begin{equation}
    \frac{\ln Z_0}{N} = \frac{\ln \left[  \mathrm{Tr} \bigl(\mathcal{I} \bigr) \right]}{N} = \ln2.
\end{equation}

The first-order ($m=1$) cumulant vanishes for isotropic interactions in zero field: $\langle \mathcal{H} \rangle_c = \langle \mathcal{H}\rangle = 0$.

The second-order ($m=2$) cumulant gives the first non-trivial correction. For a general coupling distribution $J_{ij}$, the result for spin $S = 1/2$ is
\begin{equation}
    \frac{\langle \mathcal{H}^2 \rangle_c}{N} = \frac{3}{32} \sum_{n} z_n J_n^2,
\end{equation}
where $J_n$ and $z_n$ are, respectively, the exchange constant and the coordination number per site for interaction class $n$ (with $z_0=3$, $z_1=3$, and $z_2=6$ in our model).

For higher orders ($m \ge 3$), the enumeration of connected clusters and the evaluation of their weights become increasingly involved. In this work, we used the \texttt{HTSE-code} package\cite{HTSE-code} to compute the coefficients up to order $\beta^{10}$ for the present lattice.

The raw power series has a finite radius of convergence. To extrapolate the thermal properties toward lower temperatures, we employ Padé approximants. A $P[L,M]$ Padé approximant is a rational function $P_L(\beta)/Q_M(\beta)$, where $P_L$ and $Q_M$ are polynomials of degrees $L$ and $M$, respectively. The coefficients of these polynomials are determined by requiring that the Taylor expansion of the Padé approximant matches the known HTSE coefficients up to order $\beta^{L+M}$. In practice, we retain only admissible approximants without spurious poles in the temperature range of interest and compare them with the raw truncated series. The Padé procedure is therefore used as a controlled continuation from the high-temperature side over the temperature window relevant to the transition, rather than as an exact representation of the strict $T\to 0$ limit.

\subsection{Pad\'e Continuation and Choice of Approximants}

Suppose a thermodynamic quantity is known from HTSE as a truncated power series,
\begin{equation}
    f(\beta) = \sum_{n=0}^{n_{\max}} a_n \beta^n.
\end{equation}
A Pad\'e approximant replaces this truncated polynomial by a rational function
\begin{equation}
    f(\beta) \approx \frac{P_L(\beta)}{Q_M(\beta)} = \frac{p_0 + p_1\beta + \cdots + p_L\beta^L}{1 + q_1\beta + \cdots + q_M\beta^M},
\end{equation}
with the coefficients chosen such that the Taylor expansion of $P_L(\beta)/Q_M(\beta)$ reproduces the known series through the highest available order compatible with $L+M \le n_{\max}$. The Pad\'e construction therefore does not introduce new microscopic information; it reorganizes the same HTSE coefficients into a form that usually remains stable over a wider temperature range than the raw truncated polynomial.

Among the admissible approximants without spurious poles, the near-diagonal $P[4,4]$ form gave the smoothest continuation of the $\alpha$-phase free energy and remained consistent with the raw HTSE in the overlap regime. The entropy is more sensitive because it is obtained from a temperature derivative, and the denominator-heavy $P[4,6]$ form gave the smoothest defect-free entropy curve. These choices are not mathematically unique; they provide a controlled finite-temperature extrapolation from the high-temperature side and should be regarded as a semiquantitative description of the transition regime rather than an exact low-temperature solution.

\section{Perturbative Expansion of the Heisenberg Model}

For the $\beta$ phase, the dominant low-energy object is the strong interlayer dimer. Denoting the singlet and triplet energies of a single dimer by $E_s$ and $E_t$, respectively, its free energy is

\begin{equation}
    F_{\mathrm{dimer}}=-k_B T\ln Z_{\mathrm{dimer}}
    =-k_B T\ln\left(e^{-\beta E_s}+3e^{-\beta E_t}\right).
\end{equation}

For the shifted Heisenberg convention used in the main text, $E_s=-J_0^{\text{(inter)}}$ and $E_t=0$. This isolated-dimer expression is the simple thermodynamic form used for the final $\beta$-phase entropy and free energy in the main text.

Below we also organize the effect of the weaker inter-dimer couplings in an inverse-temperature expansion in $\beta=1/(k_B T)$. This should be understood as a high-temperature series organization of the partition function, not as an independent assumption that the full thermodynamics is controlled only by a small parameter multiplying $V$.

We consider a general Hamiltonian $H = H_0 + V$, with noncommuting pieces $H_0$ and $V$. The partition function is $Z = \mathrm{Tr} \bigl(e^{-\beta H}\bigr)$. Using the Zassenhaus formula\cite{Zassenhaus},
\begin{equation}
    e^{-\beta(H_0+V)} = e^{-\beta H_0} e^{-\beta V} e^{-\frac{\beta^{2}}{2}[H_0,V]} e^{-\frac{\beta^{3}}{6}([H_0,[H_0,V]]+2[V,[H_0,V]])}\cdots,
\end{equation}
and expanding the relevant factors to order $\beta^2$,
\begin{equation}
    e^{-\beta V} = 1 - \beta V + \frac{\beta^{2}}{2}V^{2} + O(\beta^{3}), \qquad e^{-\frac{\beta^{2}}{2}[H_0,V]} = 1 - \frac{\beta^{2}}{2}[H_0,V] + O(\beta^{4}),
\end{equation}
one obtains
\begin{equation}
    e^{-\beta(H_0+V)} = e^{-\beta H_0}\Bigl[1 - \beta V + \frac{\beta^{2}}{2}V^{2} - \frac{\beta^{2}}{2}[H_0,V]\Bigr] + O(\beta^{3}).
\end{equation}
Taking the trace gives
\begin{equation}
    Z = Z_0 - \beta\,\mathrm{Tr}\bigl(e^{-\beta H_0}V\bigr) + \frac{\beta^2}{2}\,\mathrm{Tr}\bigl(e^{-\beta H_0}V^2\bigr) - \frac{\beta^2}{2}\,\mathrm{Tr}\bigl(e^{-\beta H_0}[H_0,V]\bigr)+O(\beta^3),
\end{equation}
where $Z_0=\mathrm{Tr}\bigl(e^{-\beta H_0}\bigr)$. The commutator term vanishes by cyclicity, $\mathrm{Tr}\bigl(e^{-\beta H_0}[H_0,V]\bigr)=0$. Defining $\langle O\rangle_0=Z_0^{-1}\mathrm{Tr}\bigl(e^{-\beta H_0}O\bigr)$ and expanding the logarithm of $Z$, one obtains
\begin{equation}
    \boxed{
    \ln Z = \ln Z_0  -  \beta\langle V\rangle_0  +  \frac{\beta^{2}}{2}\Bigl(\langle V^{2}\rangle_0 - \langle V\rangle_0^{2}\Bigr)  +  O(\beta^{3})}.
\end{equation}
For isotropic systems where $\langle V\rangle_0 = 0$, this reduces to
\begin{equation}
    \ln Z = \ln Z_0 + \frac{\beta^{2}}{2}\,\langle V^{2}\rangle_0 + O(\beta^{3}).
\end{equation}

For the $\beta$ phase, we consider the effective Heisenberg Hamiltonian
\begin{equation}
    \mathcal{H} = J_0^{\text{(inter)}} \sum_{\langle i,j\rangle_{\perp},s} \mathbf{S}_i \cdot \mathbf{S}_j
            + J_1^{\text{(inter)}} \sum_{\langle i,j\rangle_{\perp},w} \mathbf{S}_i \cdot \mathbf{S}_j
            + J_2^{\text{(intra)}} \sum_{\langle i,j\rangle_{\parallel}} \mathbf{S}_i \cdot \mathbf{S}_j,
\end{equation}
where $J_0^{\text{(inter)}}$ is the strong interlayer coupling, $J_1^{\text{(inter)}}$ is the weak interlayer coupling, and $J_2^{\text{(intra)}}$ is the weak intralayer coupling.
The following expressions use this unshifted convention; the bond-constant contribution discussed above must be added when comparing absolute free energies with the shifted Hamiltonian of the main text.

We take the unperturbed part to be $H_0 = J_0^{\text{(inter)}} \sum_{\langle i,j\rangle_{\perp},s} \mathbf{S}_i \cdot \mathbf{S}_j$. Each dimer then has a singlet ground state with energy $E_s = -\frac{3}{4}J_0^{\text{(inter)}}$ and a triplet excited state with energy $E_t = \frac{1}{4}J_0^{\text{(inter)}}$. The corresponding probabilities are
\begin{equation}
    P_0 = \frac{e^{-\beta E_s}}{Z_0}, \qquad 
    P_1 = \frac{3 e^{-\beta E_t}}{Z_0}, \qquad 
    Z_0 = e^{-\beta E_s} + 3 e^{-\beta E_t}.
\end{equation}
Here $Z_0$ is the partition function of a single dimer. Since each dimer contains two spins, the corresponding contribution per spin is \(\frac{1}{2}\ln Z_0\).

The first-order term \(\langle V \rangle_0\) vanishes by symmetry for isotropic couplings. For the present structure, the second-order contribution \(\langle V^2 \rangle_0/N\) separates into two pieces,
\begin{equation}
    \frac{3}{32} \sum_{n\neq 0} z_n J_n^2
    +\frac{1}{2}\, z_2 J_2^2 \cdot 3C^{2},
\end{equation}
where the sum runs over the perturbing couplings $J_1^{\text{(inter)}}$ and $J_2^{\text{(intra)}}$. The coefficient $C$ is
\begin{equation}
    C = -\frac{1}{4}P_0 + \frac{1}{12}P_1.
\end{equation}
The second term originates from a two-bond process that simultaneously couples two dimers: one bond connects the A sites of the two dimers within the same layer through $J_2$, while the other connects their B sites. This expansion provides a consistency check for the $\beta$-phase strong-dimer picture; the final finite-temperature thermodynamics used in the main text is the isolated-dimer approximation stated at the beginning of this section.
\bibliographystyle{plain}
\bibliography{ref}